\documentclass[conference]{IEEEtran}
\usepackage{graphicx}
\usepackage{color}
\usepackage[usenames,dvipsnames,svgnames,table]{xcolor}
\usepackage{siunitx}
\usepackage{cite}
\ifCLASSINFOpdf
\else
\fi
\usepackage[cmex10]{amsmath}
\usepackage{footnote}
\makesavenoteenv{tabular}
\makesavenoteenv{table}

\usepackage[footnotesize, tight]{subfigure}

\usepackage[top=0.7in, bottom=1.1in, left=0.625in, right=0.625in]{geometry}

\DeclareRobustCommand*{\IEEEauthorrefmark}[1]{%
  \raisebox{0pt}[0pt][0pt]{\textsuperscript{\footnotesize\ensuremath{#1}}}}

\begin{document}
\title{A Real-Time Millimeter-Wave Phased Array MIMO Channel Sounder}


%

%
\author{
    \IEEEauthorblockN{C. U. Bas\IEEEauthorrefmark{1}, {\it Student Member, IEEE},
    R. Wang\IEEEauthorrefmark{1}, {\it Student Member, IEEE},
    D. Psychoudakis\IEEEauthorrefmark{2}, {\it Senior Member, IEEE}, \\
    T. Henige\IEEEauthorrefmark{2},
    R. Monroe\IEEEauthorrefmark{2}, 
    J. Park\IEEEauthorrefmark{3}, {\it Member, IEEE},
    J. Zhang\IEEEauthorrefmark{2}, {\it Fellow, IEEE}, 
    A. F. Molisch\IEEEauthorrefmark{1}, {\it Fellow, IEEE}  
    }\\
         \IEEEauthorblockA{\IEEEauthorrefmark{1}University of Southern California, Los Angeles, CA, USA,}
    \IEEEauthorblockA{\IEEEauthorrefmark{2}Samsung Research America, Richardson, TX, USA}
     \IEEEauthorblockA{\IEEEauthorrefmark{3}Samsung Electronics, Suwon, Korea}

}


\maketitle
\IEEEpeerreviewmaketitle
\begin{abstract}
In this paper, we present a novel real-time MIMO channel sounder for 28 GHz. Until now, the common practice to investigate the directional characteristics of millimeter-wave channels has been using a rotating horn antenna. The sounder presented here is capable of performing horizontal and vertical beam steering with the help of phased arrays. Thanks to the fast beam-switching capability, the proposed sounder can perform measurements that are directionally resolved both at the transmitter (TX) and receiver (RX) as fast as 1.44 milliseconds compared to the minutes or even hours required for rotating horn antenna sounders. This does not only enable us to measure more points for better statistical inference but also allows us to perform directional analysis in dynamic environments. Equally important, the short measurement time combined with the high phase stability of our setup limits the phase drift between TX and RX, enabling phase-coherent sounding of all beam pairs even when TX and RX are physically separated and have no cabled connection for synchronization. This ensures that the measurement data is suitable for high-resolution parameter extraction algorithms. Along with the system design and specifications, this paper also discusses the measurements performed for verification of the sounder. Furthermore, we present sample measurements from a channel sounding campaign performed on a residential street. 
\end{abstract}

\section{Introduction} \label{sec_intro}

The ever-growing need for higher data rates in wireless communications is motivating the use of previously unused spectrum.  Consequently, the millimeter wave (mm-wave) band has become a key area of interest for next generation wireless communication systems due to  the ample amount of bandwidth available at the frequencies higher than 6 GHz. It is now clear that mm-wave systems will be an essential component of 5th generation cellular networks \cite{rangan2014millimeter}.

The knowledge of true statistical characteristics of the propagation channel is imperative for designing and testing wireless systems \cite{Molisch_2010_book}\cite{Molisch_2016_eucap}. An accurate channel model is even more important for mm-wave bands, due to its unique propagation characteristics. It is anticipated that most of the future mm-wave systems will utilize beam-forming antenna arrays to overcome the higher path loss that occurs at higher frequencies. Hence the angular spectrum and its temporal evolution are vital for the efficient design of such systems  \cite{roh2014millimeter}.

In this paper, we present a real-time double-directional channel sounder setup that is suitable for super-resolution parameter estimation. The setup presented in this paper operates at 28 GHz, although its design principles can be applied to other mm-wave bands. Unlike the commonly used virtual array, rotating horn antenna or switched antenna setups, this sounder is based on an approach of electrically-switched beams. By using arrays of antennas with phase shifters, we form beams into different directions at both the transmitter (TX) and the receiver (RX). Thanks to the control interface implemented in FPGA, we are capable of switching from one beam to another in less than 2 $\mu s$. Most of the existing directional channel sounders for indoor mm-wave systems are based on vector network analyzers (VNAs), which cannot operate in real-time, and need cabled connections between TX and RX; they are thus mostly used for static indoor channel measurements. For outdoor measurements, the prevalent method is based on mechanically rotating horn antennas \cite{Rappaport_et_al_2013_AP} \cite{Rappaport_et_al_2015_TCom}\cite{ hur_synchronous_2014} \cite{Kim_et_al_2015} \cite{MacCartney_2017_flexible}. Note, however, that mechanical rotation requires measurement durations on the order of {\em 0.5-5 hours} for one measurement location.

Since our proposed setup decreases the measurement time for a single MIMO snapshot from hours to milliseconds, it allows collection of tens of thousands of measurement points in a single measurement campaign, i.e., many orders of magnitude faster than with rotating horns. Furthermore, since all TX and RX antenna pairs can be measured within the coherence time of the channel, the developed sounder is suitable for measurement campaigns in dynamic environments. Furthermore, with proper repetition of MIMO snapshots, the temporal evolution of multi-path components (MPC) can be tracked.

Most importantly, short measurement time together with careful RF design to reduce phase noise limits the relative accumulated phase drift between the sample clocks of the TX and RX within a single MIMO snapshot even without a cabled connection for synchronizing the clocks. Thanks to this phase coherence, the measurement data are suitable for high resolution parameter extractions algorithms such as RIMAX \cite{richter2005estimation} or SAGE \cite{fleury_1999_sagel} \cite{almers2005effect}. These algorithms allow us to reveal the true double-directional channel model which only depends on the channel of interest and exclude all the effects from the hardware used in the measurements \cite{Steinbauer_et_al_2001}.


Parallel to our work, three other groups have developed real-time capable mm-wave sounders: \cite{Papazian_et_al_2016} recently presented a sounder based on an array of horn antennas combined with an electronic switch at the RX and a single TX antenna. This design is capable of faster sounding, and is similar in spirit to our approach; however, it can only provide angular information for RX side (though to the best of our knowledge it is being upgraded to a MIMO setup). There are two important differences to our setup: (i) it has significantly less equivalent isotropically radiated power (EIRP) even with the horn antennas (ii) the use of mechanically arranged horns limits the flexibility compared to our sounder, which can reconfigure the beam-patterns through software. The sounder described in \cite{Salous_2016_EuCAP} is a multi-band and multi-antenna channel sounder operating at 30 GHz, 60 GHz and 90 GHz with 8 by 8, 2 by 2 and 2 by 2 antenna arrays, respectively. However, due to the power limitations of the switch, the nominal TX power at 30 GHz is only 16 dBm and the MIMO order is lower than the 16x16 achieved in our setup. Finally, \cite{Wen_2016_mmwave} presents a sounder with 4 TX antennas multiplexed with a switch and 4 RX antennas with 4 down-conversion chains. Similar to \cite{Salous_2016_EuCAP}, the TX power is limited to 24 dBm.

The rest of the paper is organized as follows. Section \ref{sec_design} discusses the proposed channel sounder setup.  Section \ref{sec_verification} explains the measurements which investigate system performance. Section \ref{sec_post}  presents sample results from measurements. Finally Section \ref{sec_conc} summarizes results and suggests directions for future work.

\begin{figure}[tbp]    
  \centering\includegraphics[width=0.86\linewidth]{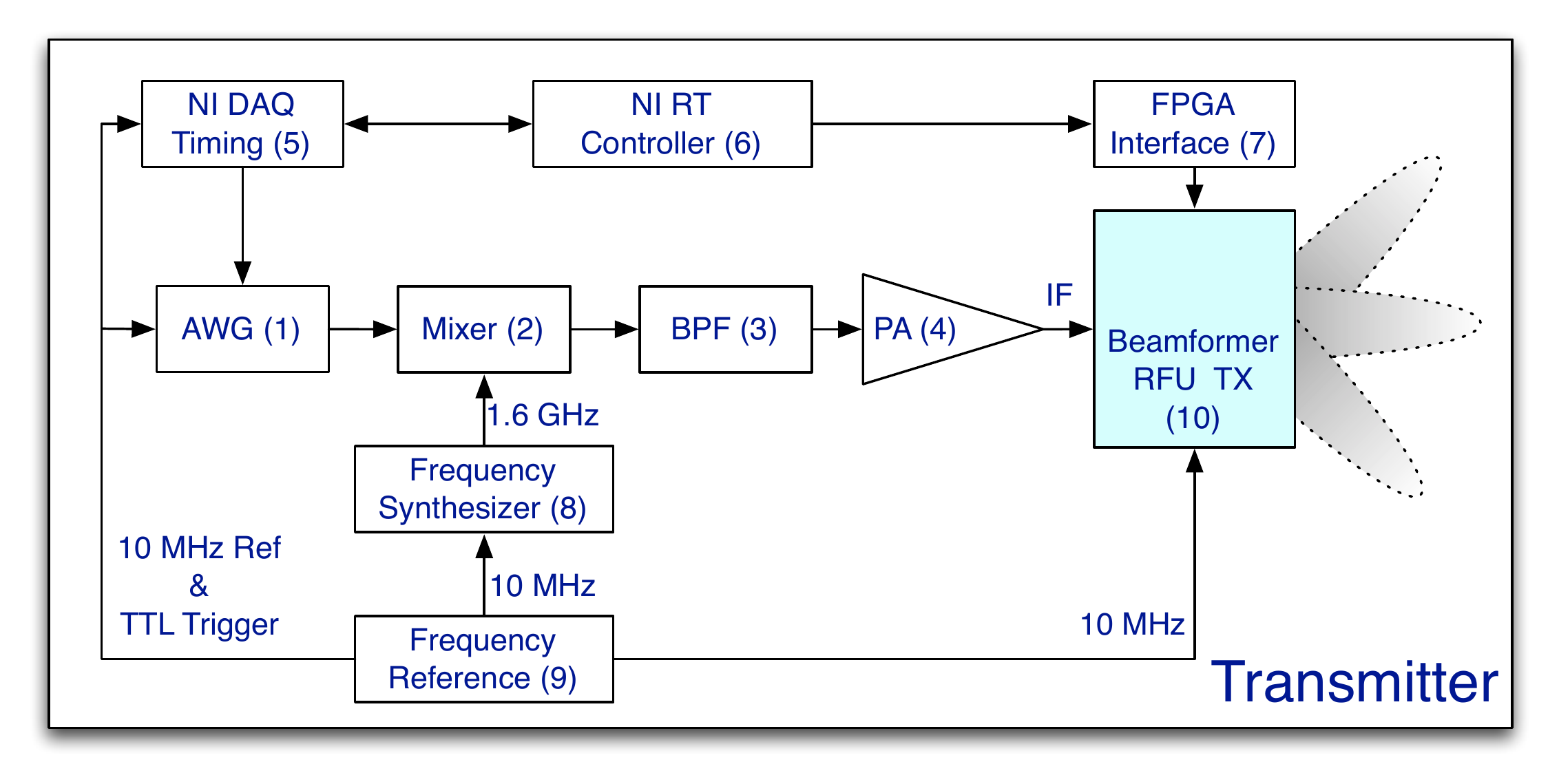}\caption{TX Block Diagram}\label{fig:TX}   
  \vspace{10pt}
  \centering\includegraphics[width=0.86\linewidth]{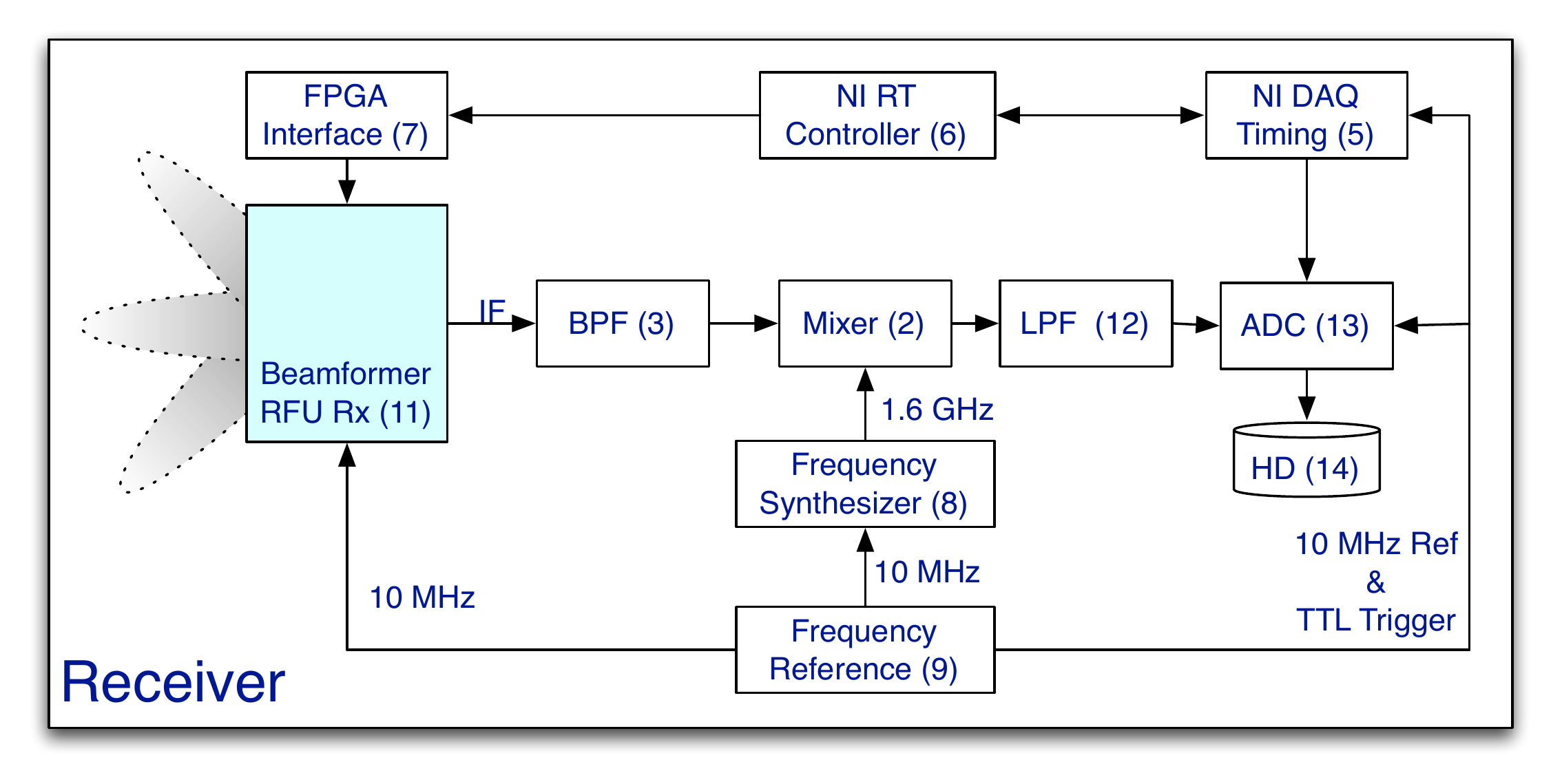}\caption{RX Block Diagram}\label{fig:RX}
\end{figure}
  \begin{table}[tbp] 
    \scriptsize
    \renewcommand{\arraystretch}{1}
      \centering     \caption{List of part numbers}
      \begin{tabular}{|c|l|}
    \hline
      (1) & Agilent N8241 \\ 
      (2) & Mini-Circuits ZEM-M2TMH+  \\ 
      (3) & KL Microwave 11ED50-1900/T500-O/O \\ 
      (4) & Pre-Amplifier \\ 
      (5) & National Instruments PXIe-6361 \\
      (6) & National Instruments PXIe-8135 \\
      (7) & National Instruments PXIe-7961R \\
      (8) & Phase Matrix FSW-0020 \\
      (9) & Precision Test Systems GPS10eR \\
      (10) & Samsung 28 GHz RFU TX \\
      (11) & Samsung 28 GHz RFU RX \\
      (12) & Low Pass filter \\
      (13) & National Instruments PXIe-5160 \\
      (14) & National Instruments HDD-8265 \\
      \hline
    \end{tabular}
  \end{table} 
\begin{figure}[tbp]
        \centering\includegraphics[width=0.85\linewidth]{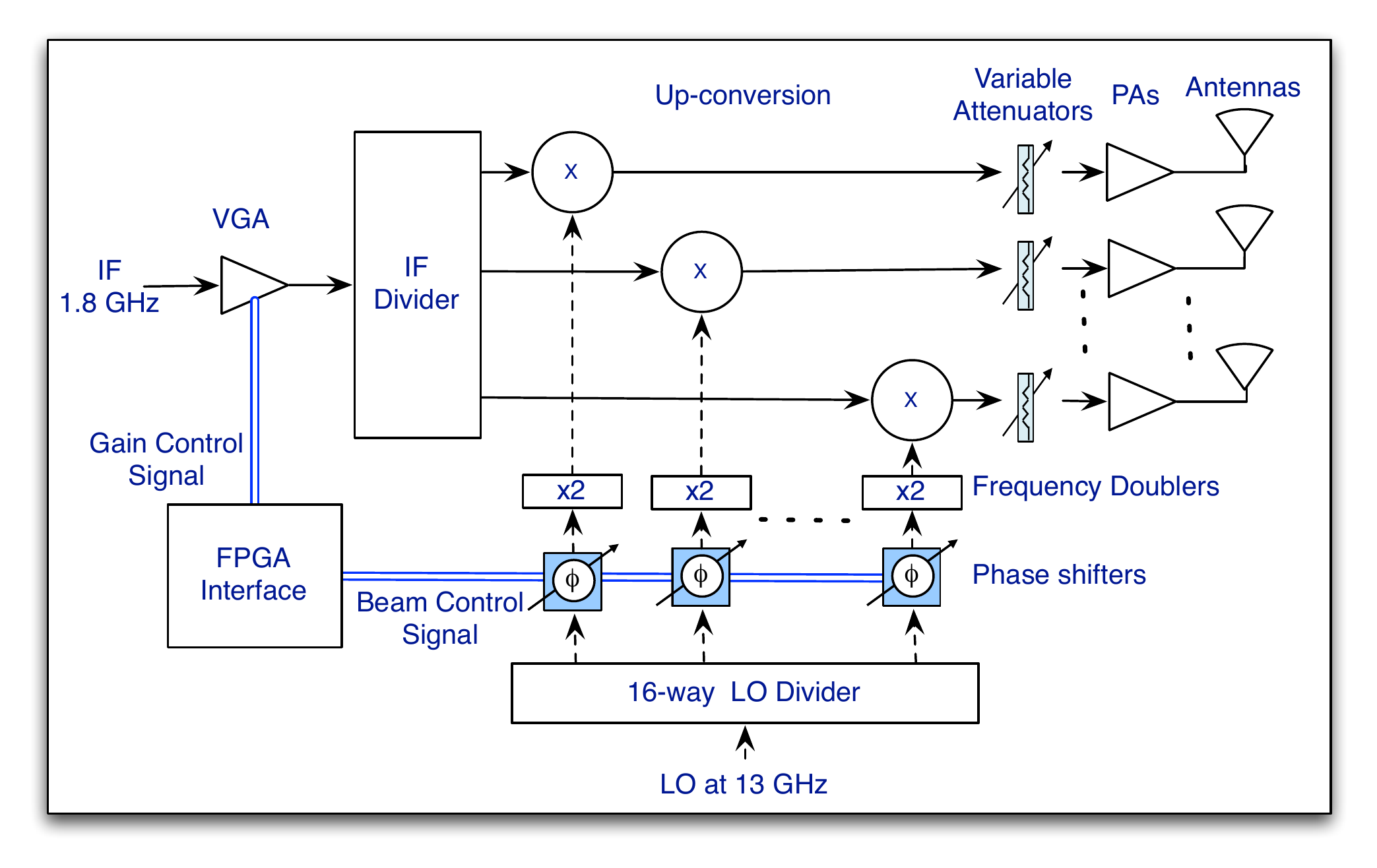}\caption{TX RFU Block Diagram}\label{fig:RFU}
\end{figure}  
  
\begin{figure}[tbp]
        \centering\includegraphics[width=0.55\linewidth, viewport=38 180 550 600, clip=true]{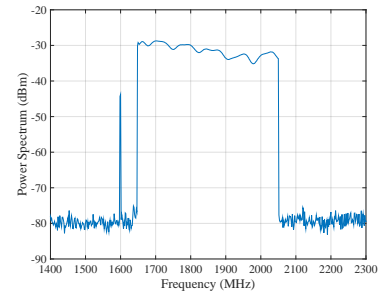}\caption{Spectrum of the sounding signal at IF}\label{fig:sounding_sig}
\end{figure}

\section{Channel Sounder Design} \label{sec_design}

The developed channel sounder is a beam-switched multi-carrier setup with 400 MHz instantaneous bandwidth. Figure \ref{fig:TX} shows the block diagram for the TX. A 15-bit, 1.25-GSps arbitrary waveform generator (AWG) generates the baseband sounding signal which consists of equally spaced 801 tones covering the frequency range from 50 MHz to 450 MHz. As suggested in \cite{Friese1997multitone}, by manipulating the phases of individual tones, we achieved a low peak to average power ratio (PAPR) of $0.4$ dB which allows us to transmit with power as close as possible to the 1 dB compression point of the power amplifiers without driving them into saturation. Since it is desirable to stay in the power amplifiers' linear regime of operation while maximizing the transmitted power, the output power is set 3.4 dB (PAPR plus 3 dB) less than the 1dB compression point of the amplifiers. Note that while Zadoff-Chu sequences, which are, e.g., used in LTE, provide PAPR=1 under idealized circumstances, this does not hold true for {\em filtered, oversampled} sequences, which are relevant here.  Our sequences outperform Zadoff-Chu by more than 1 dB. 

After the baseband signal is generated, a mixer up-converts it with a local oscillator (LO) frequency of 1.6 GHz. Since we only utilize the upper sideband of the up-converted signal as the IF input of the beam-former radio frequency unit (BF-RFU), a band-pass filter suppresses the LO leakage of the mixer and the lower sideband. After the band-pass filter and the preamplifier, we obtain an IF signal with 400 MHz bandwidth centered at 1.85 GHz as seen in Figure \ref{fig:sounding_sig}. The peak seen at the frequency of 1.6 GHz is due to residual LO leakage. It is at least 10 dB lower than the sounding tones and approximately 40 dB lower than the total power of the multi-tone signal, hence it does not affect the measurements. Finally the IF signal fed to the BF-RFU, whose simplified block diagram is shown in Figure \ref{fig:RFU}. Similar to the \cite{Psychoudakis_2016_mobile}, within the BF-RFU, this IF signal is divided into 16 RF chains. Each chain is up-converted to 27.85 GHz with dedicated amplifiers and phase shifters to perform beam-forming. The BF-RFU allows us to perform 90 degree horizontal beam steering and 60 degree vertical beam steering with 5-degree resolution. Another key feature of the RFUs is their low phase noise, see Section \ref{sec:phase}. Note that the 1 dB compression point of the individual amplifier is $31$ dBm per amplifier, however, thanks to $16$ power amplifiers used in parallel, the total output power is $40$ dBm with $3$ dB back off. After $3$ dB feed loss and taking into account the $20$ dBi antenna gain (array gain plus element gain), we achieve $57$ dBm EIRP at the output of the TX array. Combined with RX gain, for 400 MHz bandwidth, we achieve a measurable path-loss of $159$ dB without any averaging. For larger path-loss, we can use a smaller bandwidth or employ averaging at the RX.  

Similar to the TX RFU, the RX RFUs also consist of 16-element beam-forming antenna arrays. Once all received power is combined, there is an additional step of automatic gain control to utilize the dynamic range of the RX to its limits. In the current design, we have one RX BF-RFU providing 90-degree coverage. In the future, we aim to have 4 BF-RFUs for the RX to achieve complete 360 degree coverage. In the meantime, we physically rotate the RX to 4 orientations to cover 360-degree in azimuth. At the output of the RX BF-RFU, the received IF signal is filtered, down-converted back to baseband and finally sampled by a 1.25 GSps 10-bit digitizer. The digitizer streams the sampled data to a raid array with a rate up to 700 MBps and stores the data for post processing as shown in Figure \ref{fig:RX}. 

\begin{figure}[tbp]
        \centering\includegraphics[width=0.4\linewidth]{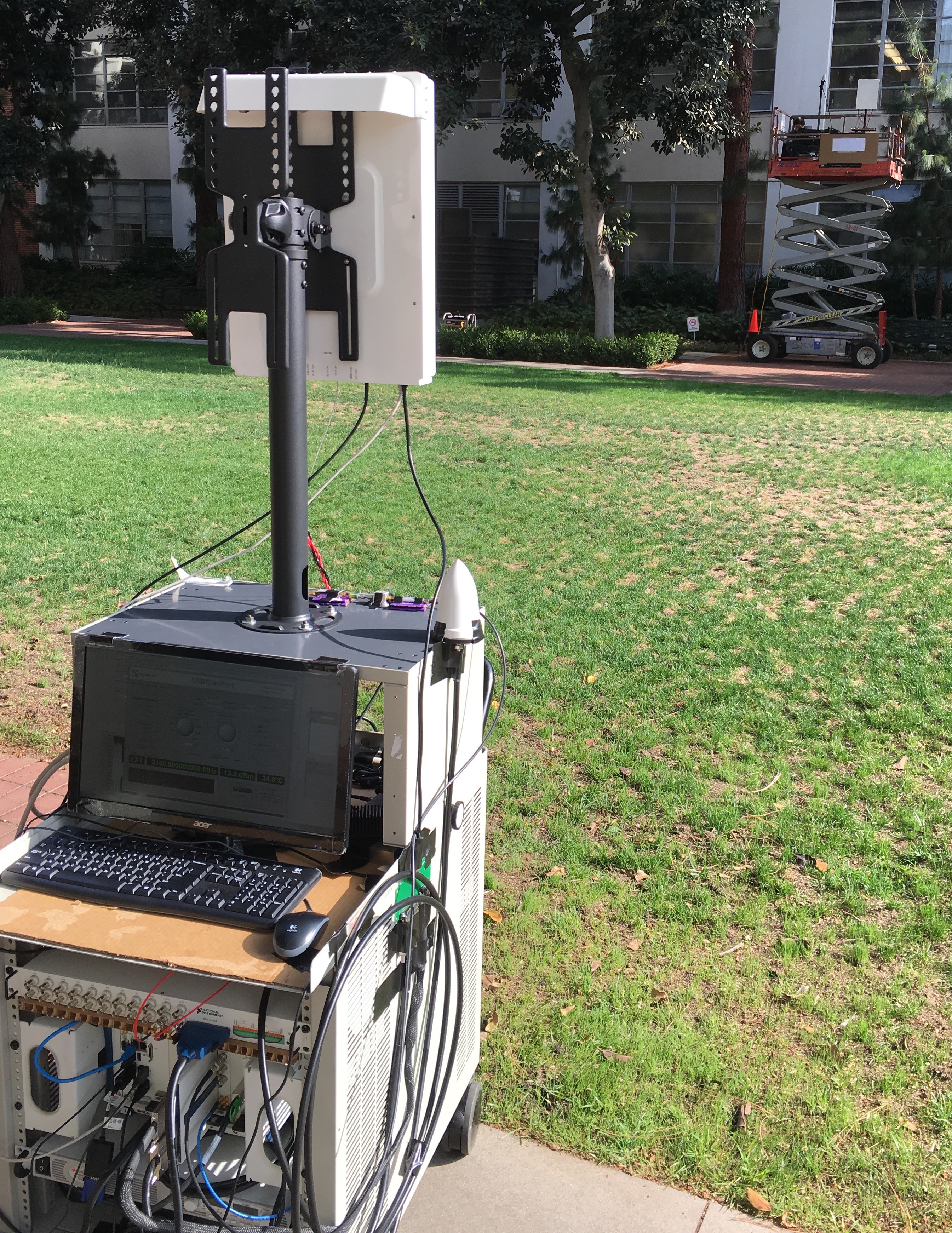}\caption{RX setup}\label{fig:rx_cart}
\end{figure}

Both the TX and the RX are controlled with LabVIEW scripts running on National Instruments PXIe controllers. The beam steering and gain of the variable gain amplifiers at the BF-RFUs are controlled via an FPGA interface with a custom designed control signaling protocol implemented in LabVIEW FPGA. This interface allows us to switch between any beam setting or gain setting in less than 2$\mu s$. Consequently, the proposed sounder can complete a full-sweep a million times faster than a virtual array. More importantly, all beam pairs can be measured without retriggering the digitizer or the AWG. This avoids triggering jitter which would create uncertainty in the absolute delay of the paths observed in different beams. TX and RX have no physical connections and they are synchronized with GPS-disciplined Rubidium frequency references. These references provide two signals for the timing of the setup; a 10 MHz clock to be used as a timebase for all units and 1 pulse per second (PPS) signal aligned to Universal Time Coordinated (UTC).

\begin{table}[tbp]\centering
      \scriptsize
  \caption{Sounder specifications}
  \renewcommand{\arraystretch}{1.1}
\begin{tabular}{l|c}
    \hline
    \multicolumn{2}{c}{Hardware Specifications} \\ \hline \hline
    Center Frequency & 27.85 GHz\\
    Instantaneous Bandwidth & 400 MHz (max 1 GHz)\\
    Antenna array size & 8 by 2 (for both TX and RX) \\
    Horizontal beam steering & -45 to 45 degree \\
    Horizontal 3dB beam width & 12 degrees\\
    Vertical beam steering & -30 to 30 degree \\
    Vertical 3dB beam width & 22 degrees\\
    Horizontal/Vertical steering steps & 5 degrees\\
    Beam switching speed & 2$\mu s$ \\
    TX EIRP & 57 dBm \\
    RX noise figure & $\le$ 5 dB \\ 
    ADC/AWG resolution & 10/15-bit \\
    Data streaming speed & 700MBps \\ \hline
    \multicolumn{2}{c}{Sounding Waveform Specifications} \\ \hline \hline
    Waveform duration & 2 $\mu s$ \\
    Repetition per beam pair & 10 \\
    Number of tones & 801 \\
    Tone spacing & 500 kHz \\
    PAPR & 0.4 dB \\ 
    Total sweep time\footnote{For 1 elevation and all 19x19 RX-TX azimuth beam combinations.} & 14.44 ms (min 1.44ms) \\ \hline  
    
  \end{tabular} \label{specs}
\end{table}
\begin{figure}[tbp]
        \centering\includegraphics[width=0.6\linewidth]{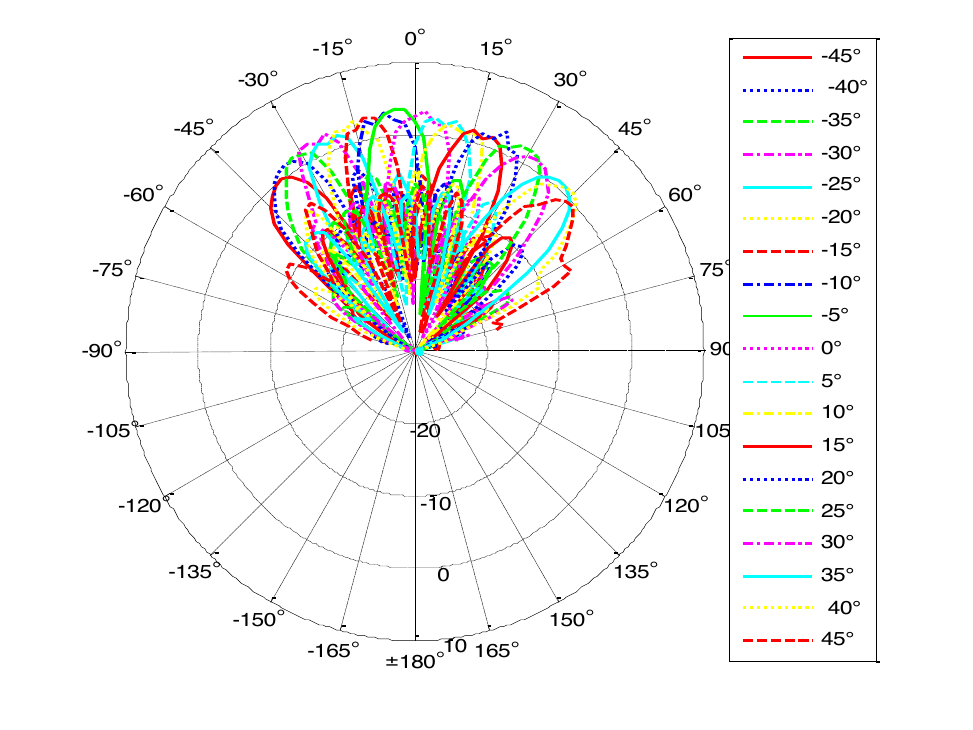}\caption{Beam patterns in Azimuth}\label{fig:pattern}
\end{figure}

\begin{figure*}[tbp]  
      \centering\includegraphics[width=0.75\linewidth, viewport=30 25 1150 345, clip=true]{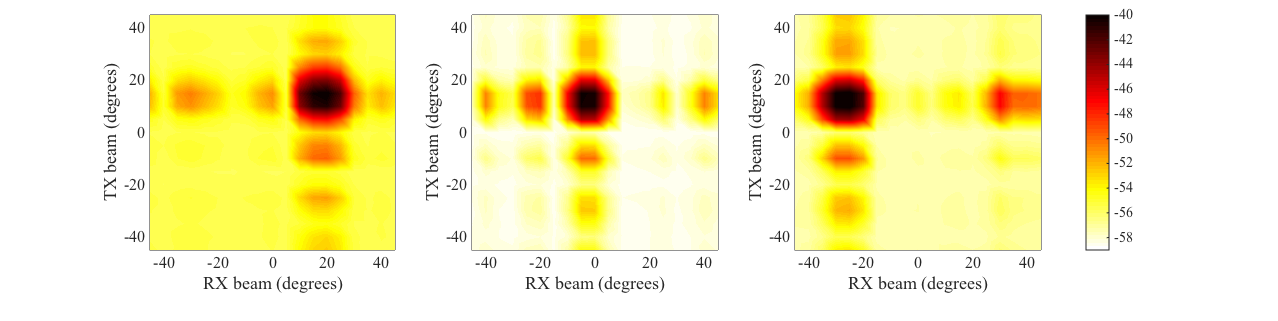}\caption{Received power vs TX and RX beam pairs when LOS component is at RX angle $x$ and TX angle $y$ and $(x,y)=\{(20,15),(-5,15),(-30,15)\}$ }\label{fig:beams}
\end{figure*}  

\begin{figure}[tbp]
        \centering\includegraphics[width=0.62\linewidth, viewport=38 180 550 600, clip=true]{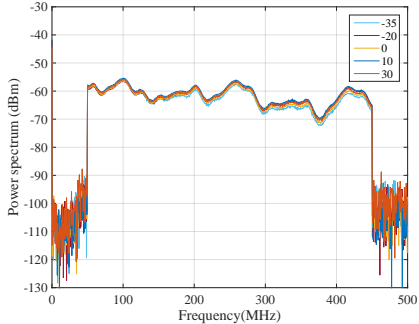}\caption{Received Power Spectrum with different RX beams }\label{fig:frequency}
\end{figure}

Given the measurement period, hardware counters in the NI DAQ Timing modules count the rising edges of the 10 MHz and trigger the rest of the units at given times. These counters are in turn triggered by the 1 PPS. Since the 1 PPS signals in the TX and RX are both aligned to the UTC, they operate synchronously without requiring any physical connections. More importantly, the AWG, the ADC, the frequency synthesizers and the BF-RFUs are disciplined with the 10 MHz signal provided by these frequency references, so that they maintain phase stability during the measurements, which is essential for accurate measurement results \cite{almers2005effect}. These references also provide GPS locations which are logged along with the measurement data.

Finally, Table \ref{specs} summarizes the hardware and sounding waveform specifications. While this is the configuration used throughout this paper, the sounding waveform can be modified without any significant changes in the hardware. Equally importantly, the beams can be configured by modifications of the FPGAs in the RFUs. This enables, for example, to get fast scanning in azimuth only for situations where we know that MPCs are mainly incident in the horizontal plane, while a slower sweep through azimuth and elevation can be implemented for other cases. This is a significant advantage compared to setups of horn arrays, which cannot be reconfigured without extensive mechanical modifications.

\begin{figure*}
\centering
\begin{minipage}{.3\textwidth}
  \centering
\includegraphics[width=0.9\linewidth, viewport=38 180 550 600, clip=true]{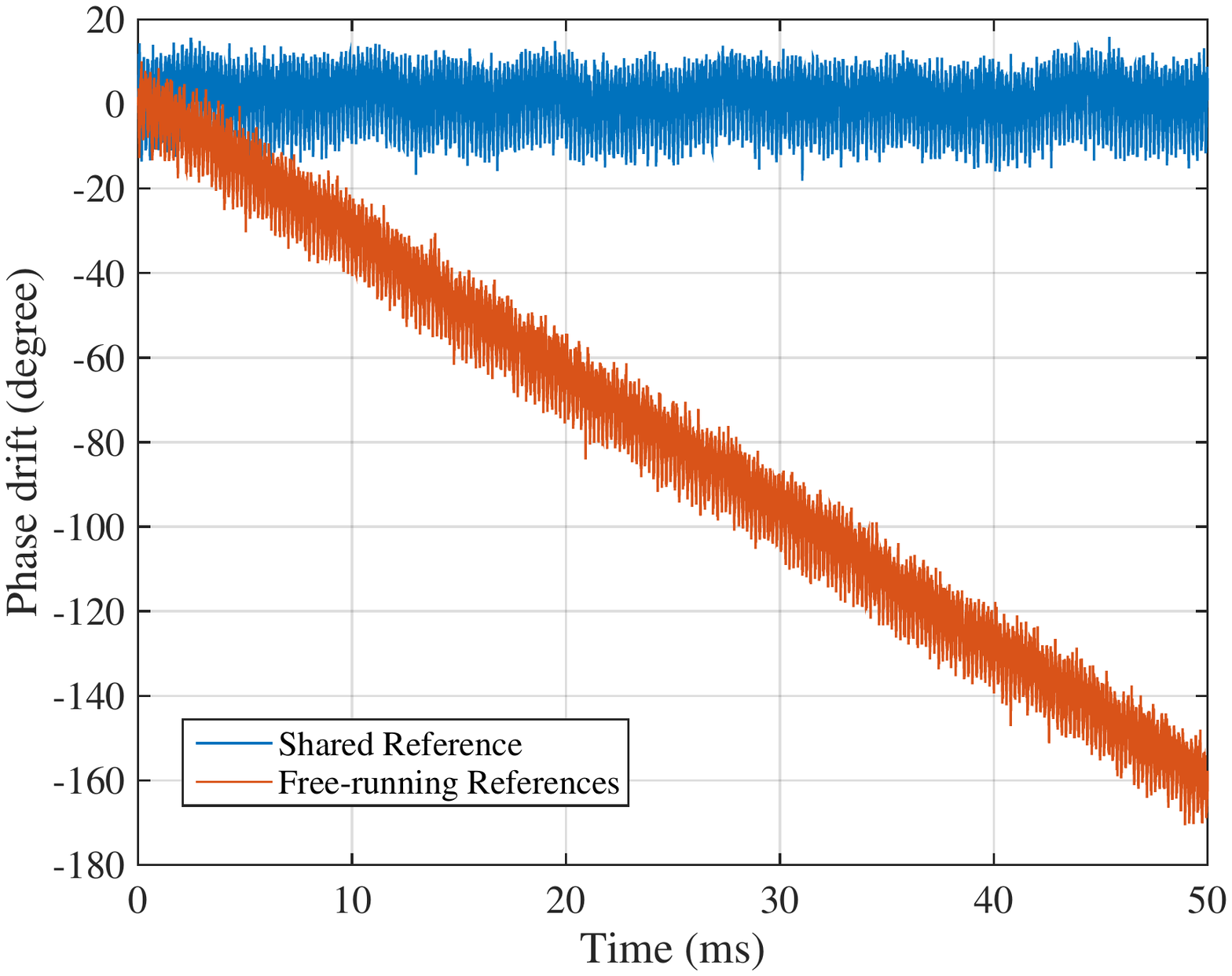}\caption{Phase drift with shared and free-running references}\label{fig:drift}
  \label{fig:test1}
\end{minipage}%
\begin{minipage}{.05\textwidth}
  
\end{minipage}
\begin{minipage}{.3\textwidth}
  \centering
\includegraphics[width=0.9\linewidth, viewport=38 180 550 600, clip=true]{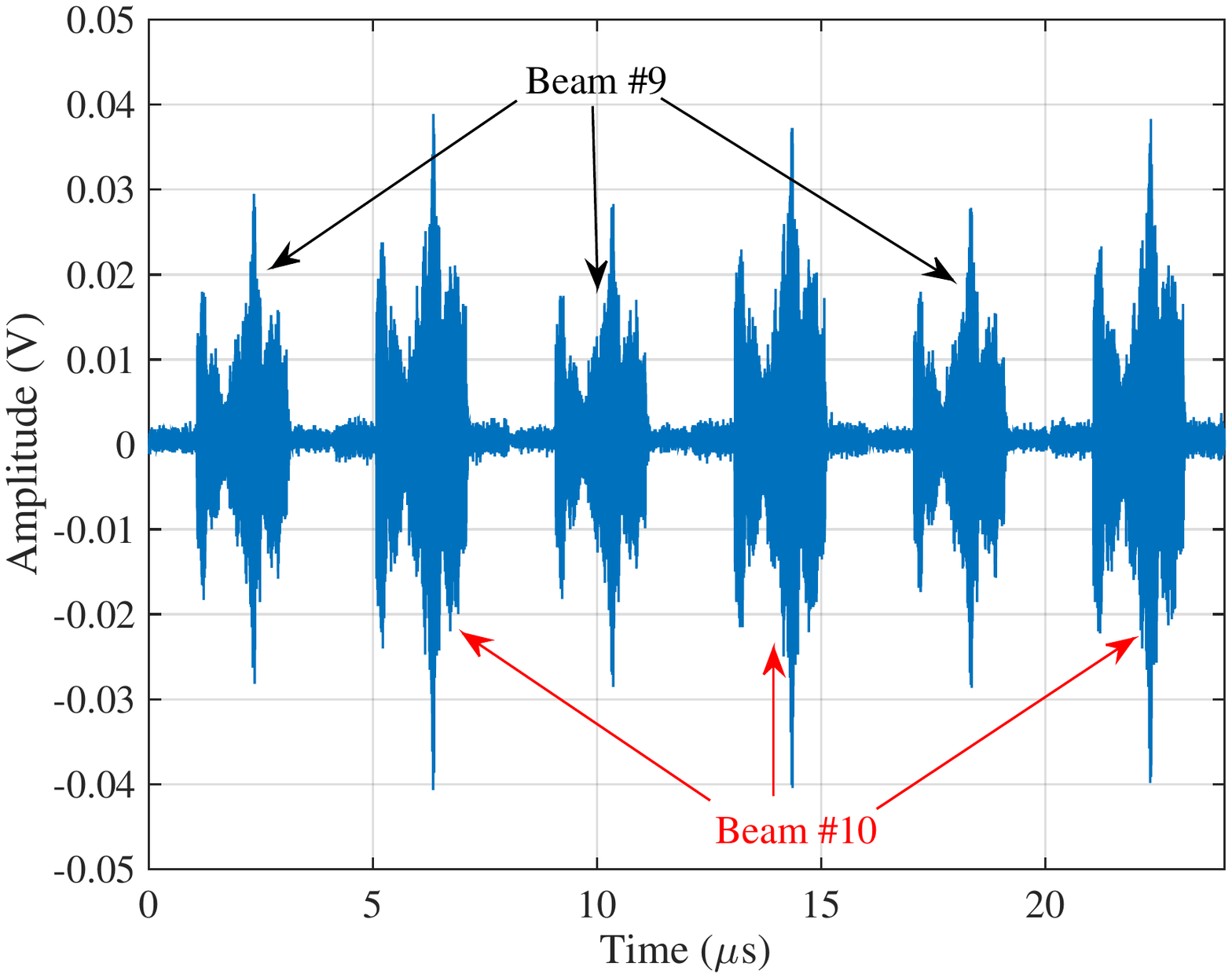}\caption{Received waveform when switching between 2 beams}\label{fig:beam_switch}
\end{minipage}
\begin{minipage}{.05\textwidth}
  
\end{minipage}
\begin{minipage}{.3\textwidth}
  \centering
\includegraphics[width=0.9\linewidth, viewport=38 180 550 600, clip=true]{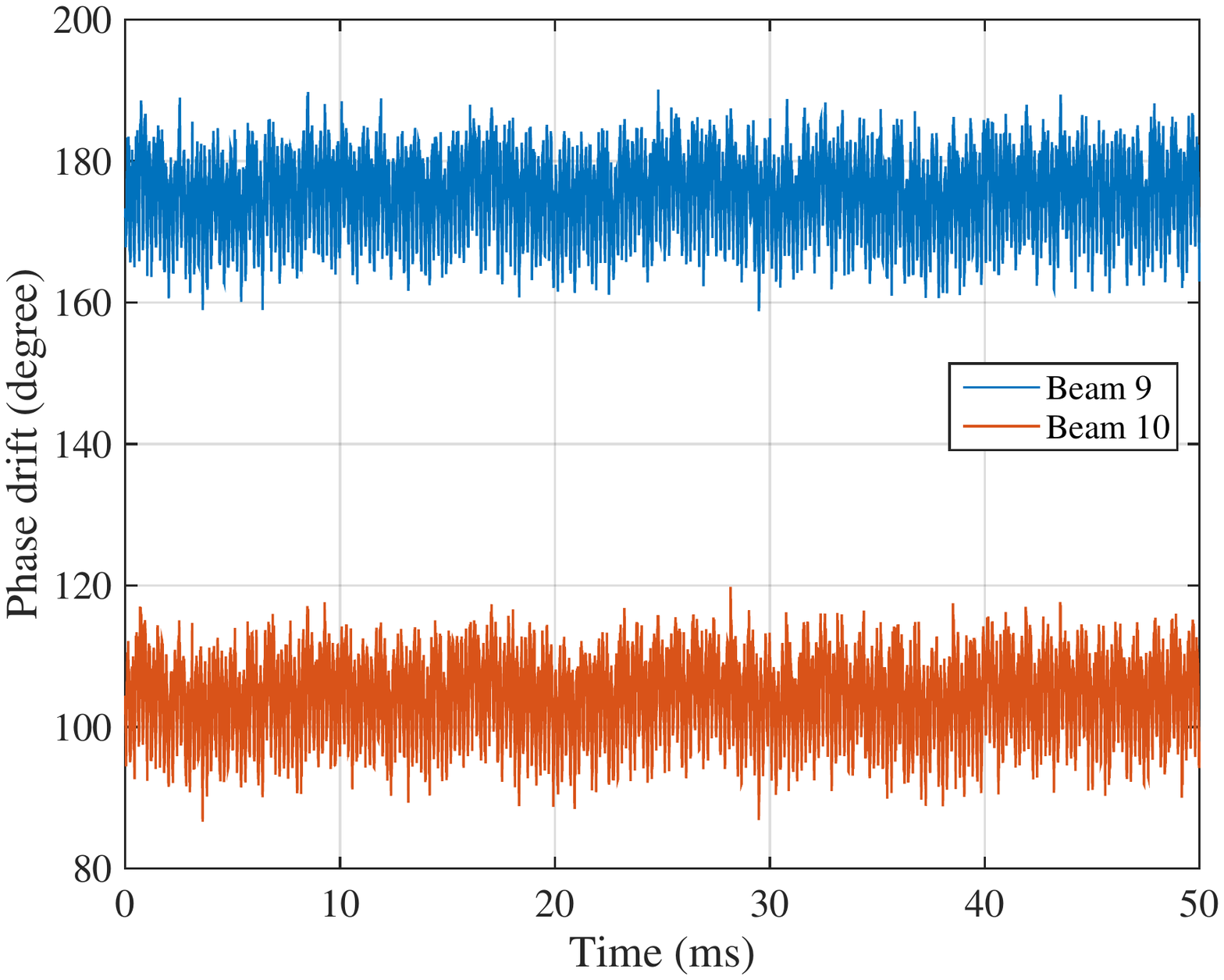}\caption{Phase drift for beam 9 and beam 10} \label{fig:beam_phases}\vspace{0.2cm}
\end{minipage}
\end{figure*}

\section{System Verification} \label{sec_verification}

\subsection{Beam Steering}

First of all, we investigate the patterns of the beams formed by the RX BF-RFU, the results are representative for the TX side as well. Figure \ref{fig:pattern} shows the beam patterns for all azimuth beams for the RX. In azimuth, 19 beams cover the range -45 to +45 degrees with 5-degree steps. All beams have 12-degree 3 dB beam-width and the side lobes are -10 dB or less relative to the main beam. Consequently, in our preliminary results, we investigate the directional channel characteristics by utilizing the beam directionality. However, with the proper calibration of the antenna arrays, in the future, we are planning to apply high resolution parameter extraction (HRPE) methods such as RIMAX to investigate directional characteristics of the propagation channel. In that case, the actual beam patterns do not affect the extracted parameters, as one can completely decouple the system response and the propagation channel \cite{richter2005estimation} \cite{Steinbauer_et_al_2001}. 

Figure \ref{fig:beams} shows the received power for all horizontal TX-RX beam pairs for where the LOS component is at x degree (relative to the 0 degree point of the physical array of the TX, and y degree relative to the RX array). Since the beams are overlapping, significant power is recorded at several beam positions. While this effect naturally limits the angular resolution when directions are determined based on the "strongest beam" only (similar to horn antennas), the effect is actually desired for HRPE algorithms which usually depend on the relative phase shift of a multi-path component received by different antennas or beams as in our case.

\subsection{Frequency Response}

Since the sounding signal has 400 MHz bandwidth, the frequency response of the channel sounder is also relevant. Figure \ref{fig:frequency} shows the combined channel response of the TX and RX. To compare the response of the different beams, we rotated the RX to align different RX beams towards to the TX. As seen in Figure \ref{fig:frequency}, the frequency responses of different RX beams are within $\pm 1$dB over the whole frequency band. Furthermore, the ratio of the complex frequency responses of two beams for all frequency tones can be estimated with a complex scalar which has approximately unit gain. This fact greatly facilitates the HRPE, see \cite{richter2005estimation}.

\subsection{Phase Stability}\label{sec:phase}

The most important features of the proposed sounder are its short measurement time and phase stability. These two features are required for measurements in dynamic environments and high resolution parameter extraction \cite{almers2005effect}. To investigate the phase stability of the system, we run the sounder continuously for 50ms with fixed beams at the TX and the RX. Figure \ref{fig:drift} shows the phase drift of the center tone for the i) best-case scenario: TX and RX use a shared reference, and ii) worst-case scenario: they operate with independent free-running Rubidium frequency references. During the measurement campaigns, the true phase stability will be in between these two cases thanks to the GPS stabilization of the frequency references. However, even with the free-running references, the total accumulated phase drift observed during a full-sweep of TX and RX beams is only 4 degrees over 1.444 ms. In this work, we employed 10 repetitions for each beam pair to improve the SNR by averaging resulting a sweep time of 14.44 ms. In this case, the accumulated drift is 40-degree in the worst case. However, as long as the total drift is guaranteed to be lower than 360 degrees, it can be estimated easily by simply adopting a switching pattern with repeated beam pairs, the estimation is further simplified by the fact that the phase drift is essentially linear even over the fairly long time of 50 ms \cite{Kristem_2017_Channel}.

Since HRPE algorithms depend on the phase relation between different beam settings, the repeatability of the phase is as important as its stability. To ensure that the calibration measurements are valid, it is crucial that the relative phase between 2 different beams stays constant at all times, even after a complete restart of the sounder. To test this, we recorded the received waveform, while switching back and forth between two beam settings as seen in Figure \ref{fig:beam_switch}. Figure \ref{fig:beam_phases} shows the phase of the center tone for the two beams. The phase offset between two beams does not change significantly over 50 ms. Furthermore, this phase offset does not change even after a complete power cycle of the sounder. Both in the frequency synthesizers and in the BF-RFUs, we have phase locked loops (PLLs) to derive the carrier frequencies from 10 MHz references. Every time they lock, they do so with a random phase, however, the effect of the random phase is the same for all beam pairs and has no effect on the {\em relative} phase offset between different beam pairs, which is the relevant quantity for HRPE.

\section{Sample Measurements} \label{sec_post}

The directional power delay profile (PDP) for the TX beam and RX beam with the azimuth angles  $\theta_{TX}$  and  $\theta_{RX}$  is estimated as;
\begin{equation}
  \resizebox{.9 \linewidth}{!} 
{
$   P(\theta_{TX},\theta_{RX},\tau) = \bigg\vert \mathcal{F}^{-1} \left\{ H_{\theta_{TX},\theta_{RX}}\left(\vec{f} \right) ./ H_{cal}\left(\vec{f} \right) \right\} \bigg\vert ^2$
}
\end{equation}
where $\theta_{TX/RX}\in[-45,45]$, $\mathcal{F}^{-1}$ denotes inverse Fourier transform, $H_{i,j}(\vec{f})$ and $H_{cal}(\vec{f})$ are the frequency responses for $i$-th TX and $j$-th RX beam and the calibration response respectively; $\vec{f}$ are the used frequency tones, and $./$ is element-wise division. 

\begin{figure}[tbp]
        \centering\includegraphics[width=0.7\linewidth,viewport=25 0 620 400, clip=true]{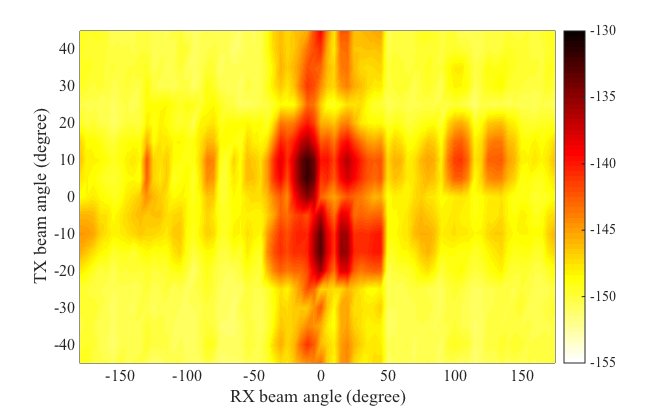}\caption{Received power vs TX and RX beams}\label{fig:rx_power}
\end{figure}

Since our current setup measures 90-degree sectors, we simply repeat the measurements when the RX is rotated to $(0,90,180,270)$ degrees to obtain 360-degree coverage at the RX side. Then the angular power spectrum can be simply calculated as:
\begin{equation}
 {\displaystyle  PAS(\theta_{TX},\theta_{RX}+\phi) = \sum_{\tau} P_\phi(\theta_{TX},\theta_{RX},\tau)}
\end{equation}
where $P_{\phi}(\cdot)$ is the directional PDP in the RX orientation $\phi \in (0,90,280,270)$. Figure \ref{fig:rx_power} shows the power angular spectrum for a sample non-line-of-sight (NLOS) measurement in a residential environment with a distance of 150 m. 

\begin{figure}[tbp]
        \centering\includegraphics[width=0.60\linewidth, viewport=38 180 550 600, clip=true]{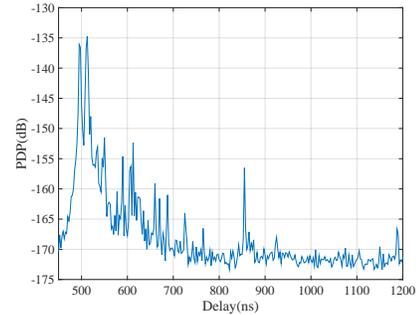}\caption{Power delay profile}\label{fig:pdp}
\end{figure}

For the 90-degree sectors in the TX and the RX, since all beam pairs are measured without a significant phase drift or trigger jitter, all directional PDPs are already aligned in the delay domain and require no further correction. Consequently, the corresponding PDP can be obtained similarly to the virtual horn measurements with a shared reference even though the sounder does not have a shared reference. There are different methods discussed in \cite{hur_synchronous_2014}  \cite{Haneda_2016_omni} for synthesizing omni-directional PDP from directional measurements. Ultimately, the PDP obtained from HRPE will provide the best representation of the channel. However, we use the approach from \cite{hur_synchronous_2014} to present sample results here. Hence, the PDP for the 90-sector with the RX orientation $\phi$ is given by:
\begin{equation}
  PDP_{\phi} (\tau) = {\displaystyle \max_{\theta_{TX}} \max_{\theta_{RX}}  P_\phi(\theta_{TX},\theta_{RX},\tau)} \label{eq:pdp}
\end{equation}

\begin{figure}[tbp]
        \centering\includegraphics[width=0.70\linewidth,viewport=10 5 525 400, clip=true]{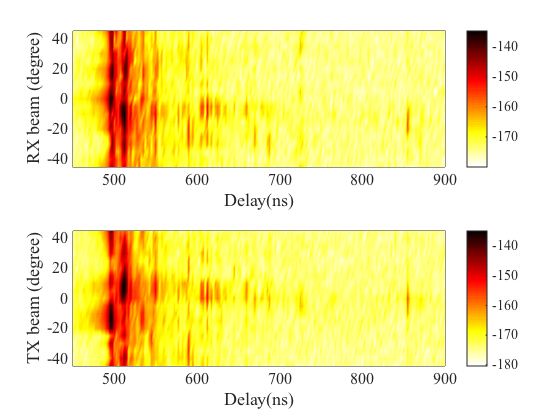}\caption{Power angular-delay profile}\label{fig:padp}
\end{figure}

Figure \ref{fig:pdp} shows the $P_{0} (\tau)$ for the same NLOS measurement location. Furthermore, Figure \ref{fig:padp} shows the power angular-delay profiles for RX and TX which are calculated as follows. 
\begin{equation} 
\begin{aligned}  
  PADP_{\phi, RX}(\theta_{RX},\tau) &= {\displaystyle \max_{\theta_{TX}} P_{\phi}(\theta_{TX},\theta_{RX},\tau)} \\
  PADP_{\phi, TX}(\theta_{TX},\tau) &= {\displaystyle \max_{\theta_{RX}} P_{\phi}(\theta_{TX},\theta_{RX},\tau)}
 \end{aligned} 
\end{equation}

\begin{figure}[tbp]
        \centering\includegraphics[width=0.60\linewidth,viewport=38 180 550 600, clip=true]{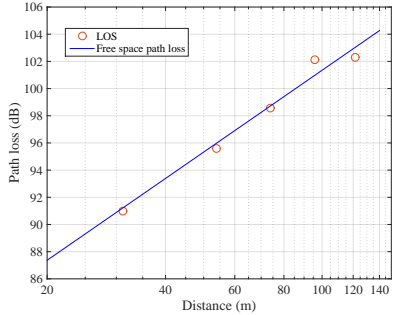}\caption{Line-of-sight (LOS) path-loss}\label{fig:los_pl}
\end{figure}

Since we have to retrigger for every orientation (90 degree steps) of the RX, the $PDP_{\phi} (\tau)$'s obtained for different $\phi$'s might have a phase offset. Consequently, synthesizing omnidirectional PDP requires more post-processing, which will be investigated in future. However, once the RX setup is upgraded for 360 coverage, there won't be any phase offset between different sectors and the omnidirectional PDP can be synthesized similar to the Equation \ref{eq:pdp}. Finally, we can obtain 360-degree path-loss by adding up the power received from all RX orientations: 

\begin{equation}
 {\displaystyle  PL =  \sum_{\phi} \sum_{\tau} PDP_\phi(\tau)} \label{eq:pl}
\end{equation}

Figure \ref{fig:los_pl} shows the calculated path-loss for line-of-sight measurements according to Equation \ref{eq:pl}. The path-loss values are in good agreement with the theoretical free-space path-loss curve also shown in the figure.

\section{Conclusion}\label{sec_conc}
In this paper, we presented a novel mm-wave channel sounder which can operate in dynamic environments. By using a beam-forming array, we managed to decrease the measurement time from minutes to milliseconds compared to rotating horn antenna approach. Furthermore, thanks to the beam-forming gain, the measurable path-loss we achieved for a 400 MHz bandwidth is 159 dB without employing waveform averaging. We also validated the sounders phase stability which is the paramount feature of the proposed design. In the future, we will upgrade our RX to cover 360 degrees in azimuth, perform extensive channel sounding campaigns and process the acquired data with high resolution parameter extraction algorithms.

\section*{Acknowledgements}
Part of this work was supported by grants from the National Science Foundation.

\bibliographystyle{IEEEtran}
\bibliography{mmwave,60ghz_models}

\begin{thebibliography}{10}
\providecommand{\url}[1]{#1}
\csname url@samestyle\endcsname
\providecommand{\newblock}{\relax}
\providecommand{\bibinfo}[2]{#2}
\providecommand{\BIBentrySTDinterwordspacing}{\spaceskip=0pt\relax}
\providecommand{\BIBentryALTinterwordstretchfactor}{4}
\providecommand{\BIBentryALTinterwordspacing}{\spaceskip=\fontdimen2\font plus
\BIBentryALTinterwordstretchfactor\fontdimen3\font minus
  \fontdimen4\font\relax}
\providecommand{\BIBforeignlanguage}[2]{{%
\expandafter\ifx\csname l@#1\endcsname\relax
\typeout{** WARNING: IEEEtran.bst: No hyphenation pattern has been}%
\typeout{** loaded for the language `#1'. Using the pattern for}%
\typeout{** the default language instead.}%
\else
\language=\csname l@#1\endcsname
\fi
#2}}
\providecommand{\BIBdecl}{\relax}
\BIBdecl

\bibitem{rangan2014millimeter}
S.~Rangan, T.~S. Rappaport, and E.~Erkip, ``Millimeter-wave cellular wireless
  networks: Potentials and challenges,'' \emph{Proceedings of the IEEE}, vol.
  102, no.~3, pp. 366--385, March 2014.

\bibitem{Molisch_2010_book}
A.~F. Molisch, \emph{Wireless communications}, 2nd~ed.\hskip 1em plus 0.5em
  minus 0.4em\relax IEEE Press - Wiley, 2010.

\bibitem{Molisch_2016_eucap}
A.~F. Molisch, A.~Karttunen, R.~Wang, C.~U. Bas, S.~Hur, J.~Park, and J.~Zhang,
  ``Millimeter-wave channels in urban environments,'' in \emph{2016 10th
  European Conference on Antennas and Propagation (EuCAP)}, April 2016, pp.
  1--5.

\bibitem{roh2014millimeter}
W.~Roh, J.~Y. Seol, J.~Park, B.~Lee, J.~Lee, Y.~Kim, J.~Cho, K.~Cheun, and
  F.~Aryanfar, ``{Millimeter-wave beamforming as an enabling technology for 5G
  cellular communications: theoretical feasibility and prototype results},''
  \emph{IEEE Communications Magazine}, vol.~52, no.~2, pp. 106--113, February
  2014.

\bibitem{Rappaport_et_al_2013_AP}
T.~S. Rappaport, F.~Gutierrez, E.~Ben-Dor, J.~N. Murdock, Y.~Qiao, and J.~I.
  Tamir, ``Broadband millimeter-wave propagation measurements and models using
  adaptive-beam antennas for outdoor urban cellular communications,''
  \emph{Antennas and Propagation, IEEE Transactions on}, vol.~61, no.~4, pp.
  1850--1859, April 2013.

\bibitem{Rappaport_et_al_2015_TCom}
T.~S. Rappaport, G.~R. Maccartney, M.~K. Samimi, and S.~Sun, ``Wideband
  millimeter-wave propagation measurements and channel models for future
  wireless communication system design,'' \emph{Communications, IEEE
  Transactions on}, vol.~63, no.~9, pp. 3029--3056, Sept 2015.

\bibitem{hur_synchronous_2014}
S.~Hur, Y.~J. Cho, J.~Lee, N.-G. Kang, J.~Park, and H.~Benn, ``{Synchronous
  channel sounder using horn antenna and indoor measurements on 28 GHz},'' in
  \emph{2014 IEEE International Black Sea Conference on Communications and
  Networking (BlackSeaCom)}, May 2014, pp. 83--87.

\bibitem{Kim_et_al_2015}
M.-D. Kim, J.~Liang, Y.~K. Yoon, and J.~H. Kim, ``{28 GHz path loss
  measurements in urban environments using wideband channel sounder},'' in
  \emph{Antennas and Propagation USNC/URSI National Radio Science Meeting, 2015
  IEEE International Symposium on}, July 2015, pp. 1798--1799.

\bibitem{MacCartney_2017_flexible}
G.~R. MacCartney and T.~S. Rappaport, ``A flexible millimeter-wave channel
  sounder with absolute timing,'' \emph{IEEE Journal on Selected Areas in
  Communications}, vol.~PP, no.~99, pp. 1--1, 2017.

\bibitem{richter2005estimation}
A.~Richter, ``Estimation of radio channel parameters: Models and
  algorithms.''\hskip 1em plus 0.5em minus 0.4em\relax ISLE, 2005.

\bibitem{fleury_1999_sagel}
B.~H. Fleury, M.~Tschudin, R.~Heddergott, D.~Dahlhaus, and K.~I. Pedersen,
  ``{Channel parameter estimation in mobile radio environments using the SAGE
  algorithm},'' \emph{IEEE Journal on selected areas in communications},
  vol.~17, no.~3, pp. 434--450, Mar 1999.

\bibitem{almers2005effect}
P.~Almers, S.~Wyne, F.~Tufvesson, and A.~F. Molisch, ``{Effect of random walk
  phase noise on MIMO measurements},'' in \emph{Vehicular Technology
  Conference, 2005. VTC 2005-Spring. 2005 IEEE 61st}, vol.~1.\hskip 1em plus
  0.5em minus 0.4em\relax IEEE, May 2005, pp. 141--145.

\bibitem{Steinbauer_et_al_2001}
M.~Steinbauer, A.~F. Molisch, and E.~Bonek, ``{The double-directional radio
  channel},'' \emph{Antennas and Propagation Magazine, IEEE}, vol.~43, no.~4,
  pp. 51--63, Aug 2001.

\bibitem{Papazian_et_al_2016}
P.~B. Papazian, C.~Gentile, K.~A. Remley, J.~Senic, and N.~Golmie, ``A radio
  channel sounder for mobile millimeter-wave communications: System
  implementation and measurement assessment,'' \emph{IEEE Transactions on
  Microwave Theory and Techniques}, vol.~64, no.~9, pp. 2924--2932, Sept 2016.

\bibitem{Salous_2016_EuCAP}
S.~Salous, ``{Multi-band multi-antenna chirp channel sounder for frequencies
  above 6 GHz},'' in \emph{2016 10th European Conference on Antennas and
  Propagation (EuCAP)}, April 2016, pp. 1--4.

\bibitem{Wen_2016_mmwave}
Z.~Wen, H.~Kong, Q.~Wang, S.~Li, X.~Zhao, M.~Wang, and S.~Sun, ``{mmWave
  channel sounder based on COTS instruments for 5G and indoor channel
  measurement},'' in \emph{2016 IEEE Wireless Communications and Networking
  Conference Workshops (WCNCW)}, April 2016, pp. 37--43.

\bibitem{Friese1997multitone}
M.~Friese, ``Multitone signals with low crest factor,'' \emph{Communications,
  IEEE Transactions on}, vol.~45, no.~10, pp. 1338--1344, Oct 1997.

\bibitem{Psychoudakis_2016_mobile}
D.~Psychoudakis, H.~Zhou, B.~Biglarbegian, T.~Henige, and F.~Aryanfar,
  ``{Mobile station radio frequency unit for 5G communications at 28 GHz},'' in
  \emph{2016 IEEE MTT-S International Microwave Symposium (IMS)}, May 2016, pp.
  1--3.

\bibitem{Kristem_2017_Channel}
V.~Kristem, O.~Sangodoyin, C.~Bas, M.~Kaske, J.~Lee, C.~Schneider,
  G.~Sommerkorn, J.~Zhang, R.~S. Thoma, and A.~Molisch, ``{3D MIMO
  Outdoor-to-Indoor Propagation Channel Measurement},'' \emph{IEEE Transactions
  on Wireless Communications}, vol.~PP, no.~99, pp. 1--1, 2017.

\bibitem{Haneda_2016_omni}
K.~Haneda, S.~L.~H. Nguyen, J.~JŠrvelŠinen, and J.~Putkonen, ``Estimating the
  omni-directional pathloss from directional channel sounding,'' in \emph{2016
  10th European Conference on Antennas and Propagation (EuCAP)}, April 2016,
  pp. 1--5.

\end{thebibliography}

\end{document}